# Towards Understanding Worldwide Cross-cultural Differences in Implicit Driving Cues: Review, Comparative Analysis, and Research Roadmap

Yongqi Dong, Chang Liu, Yiyun Wang, and Zhe Fu

*Abstract*—Recognizing and understanding implicit driving cues across diverse cultures is imperative for fostering safe and efficient global transportation systems, particularly when training new immigrants holding driving licenses from culturally disparate countries. Additionally, it is essential to consider cross-cultural differences in the development of Automated Driving features tailored to different countries. Previous piloting studies have compared and analyzed cross-cultural differences in selected implicit driving cues, but they typically examine only limited countries. However, a comprehensive worldwide comparison and analysis are lacking. This study conducts a thorough review of existing literature, online blogs, and expert insights from diverse countries to investigate cross-cultural disparities in driving behaviors, specifically focusing on implicit cues such as non-verbal communication (e.g., hand gestures, signal lighting, honking), norms, and social expectations. Through comparative analysis, variations in driving cues are illuminated across different cultural contexts. Based on the findings and identified gaps, a research roadmap is proposed for future research to further explore and address these differences, aiming to enhance intercultural communication, improve road safety, and increase transportation efficiency on a global scale. This paper presents the pioneering work towards a comprehensive understanding of the implicit driving cues across cultures. Moreover, this understanding will inform the development of automated driving systems tailored to different countries considering cross-cultural differences.

## I. INTRODUCTION

Driving is a social task that involves both explicit and implicit interactions through driving cues. These cues exhibit significant variation across different cultural backgrounds, spanning various countries and even within different regions of the same country. For instance, differences in horn usage and lane discipline exemplify this variation.

(1) Horn usage: In the United States and Canada (except for metropolises like New York City), honking is primarily reserved for emergency situations or to alert of imminent danger. Conversely, in countries such as China, India, and Vietnam, honking serves multiple communication purposes, including expressing frustration or conveying greetings.

(2) Lane discipline: Drivers in the United States and Australia typically maintain considerable distances between vehicles and rigorously adhere to lane markings. In contrast, Italy and Greece, characterized by more communal social interactions, witness a more flexible lane discipline with frequent lane changes and closer driving proximity.

Furthermore, illustrating regional disparities within the same country, the approach to traffic signals across the United States presents notable contrasts: In bustling urban locales like New York City, drivers often exhibit assertive driving behaviors, promptly accelerating upon green lights and utilizing honking to convey impatience. Whereas, in tranquil rural regions such as Vermont, a more laid-back driving demeanor prevails. There, drivers patiently await a brief moment after a light changes to green before proceeding, and horn usage is sparse or nonexistent owing to the serene and less congested road conditions.

Understanding the implicit driving cues across cultures is crucial for promoting safe and effective transportation systems globally, especially when training new immigrants who already possess a driving license from a country with different cultural backgrounds. Additionally, regarding merging technologies, it is essential to consider cross-cultural differences in the development of Automated Driving features tailored to different countries [1]–[3]. While there have been piloting studies comparing and analyzing cross-cultural differences in implicit driving cues, e.g., [3]–[7], these studies have often been limited in scope, typically focusing on a limited number of fewer than ten countries. For instance, research has compared British and Malaysian drivers in their use of explicit (turn signals) and implicit (e.g., vehicle position, speed) communicative cues when discerning the intentions of other road users [4]. Similarly, studies have examined German and British drivers' implicit communication on motorway slip roads through longitudinal driving dynamics (acceleration, deceleration, maintaining speed) [3], as well as differences between Swedish and Turkish drivers in their intention to comply with speed limits [5]. Additionally, research has explored the driving tendencies of drivers from Finland, Sweden, Greece, and Turkey, regarding various aberrant driving behaviors [6]. Furthermore, a study has investigated various factors related to legislation, enforcement, and education aimed at fostering safer drivers across nine countries, including the United Kingdom (particularly Great Britain), Germany, Italy, Egypt, Qatar, the United Arab Emirates, China, Japan, and Canada [7].

*Research supported by the Transport & Mobility Institute at Delft University of Technology.

Yongqi Dong is with Delft University of Technology, Delft, 2628 CN, the Netherlands. (e-mail: y.dong-4@tudelft.nl).

Chang Liu with the Machine Perception Research Laboratory of Institute for Computer Science and Control (HUN-REN SZTAKI), 1117 Budapest, Hungary (corresponding author; e-mail: changliu@sztaki.hu).

Yiyun Wang is with Delft University of Technology, Delft, 2628 CN, the Netherlands. (e-mail: y.wang-31@tudelft.nl ).

Zhe Fu is with the University of California, Berkeley, CA 94720, United States. (e-mail: zhefu@berkeley.edu)

Despite the valuable insights from existing studies, there is still a notable gap in achieving a comprehensive, global perspective that integrates these findings and provides a nuanced understanding of implicit driving cues across diverse cultural contexts. To address this gap, this study conducts a thorough review of existing literature, online blogs, and expert insights from various countries to investigate cross-cultural disparities in driving behaviors. Specifically, the study focuses on implicit cues such as non-verbal communication (e.g., hand gestures, signal lighting, honking speed control and choice).

Through comparative analysis, the study sheds light on the variations in driving cues observed across different cultural contexts. Building on these findings, the study proposes a research roadmap for future investigations aimed at further exploring and addressing worldwide cross-cultural differences. The ultimate goal of this research endeavor is to enhance intercultural communication, improve road safety, and increase transportation efficiency on a global scale. By presenting pioneering work towards achieving a comprehensive understanding of implicit driving cues across cultures, this paper seeks to make a valuable contribution to the advancement of knowledge in this field. Furthermore, the insights gained from this study will be instrumental in informing the development of automated driving systems tailored to different countries, taking into consideration their unique cross-cultural dynamics influencing driving behaviors.

In the subsequent sections, this paper will commence by summarizing and comparing cross-cultural differences in implicit driving cues, categorized by the types of communication or interaction employed. Specifically, Section II will delve into lighting signals, Section III will focus on gestures, Section IV will explore honking, and Section V will examine speed control and choice. Then, Section VI will present the proposed roadmap for future research. Finally, Section VII will provide the conclusion.

## II. SIGNAL LIGHTING

Driving behaviors are often influenced by the use of lighting signals, which serve as an important means of communication on the road. However, the interpretation and significance of these signals can vary significantly across different cultures, reflecting diverse societal norms and expectations [8].

### A. Cultural Variance in Double Flashing Behavior

Car light signals, such as double flashing, flashing high beams, and the use of turn signals, are not universal in their interpretation. Instead, they reflect the diverse cultural norms and practices prevalent in different regions worldwide. For instance, in some areas, double flashing lights may convey expressions of gratitude or apology, while in others, they serve as a warning or alert to other drivers. Similarly, the act of flashing high beams or using turn signals can carry varying meanings across cultures. Understanding these cultural nuances is crucial for effective communication on the road. Therefore, the following Table I highlights the diverse interpretations of double flashing light behavior across different countries. Understanding these cultural differences can enhance communication and promote safer driving practices on the road.

TABLE I. DIFFERENT MEANINGS OF DOUBLE FLASHING BEHAVIOR

| Country | Meaning of Double Flashing Behavior |
|---|---|
| Hungary | Two quick flashes indicate gratitude or yielding. |
| Russia | Three quick flashes indicate gratitude. |
| Italy | Rapid double flashes serve as a warning to other drivers. |
| Germany | Double flashing serves as a reminder, indicating intent to overtake or merge into another lane. |
| France | Double flashes are sometimes used to mean "thank you" or "sorry", but not very common. |
| UK | Double flashes are usually used as a warning, for example, alerting other vehicles that your vehicle has a problem or needs to stop urgently. |
| China | Double flashing is often used as a warning, such as alerting of obstacles ahead. |
| Japan | Double flashes are often used to express apology or gratitude, also may be used as a reminder. |
| USA | Double flashes are sometimes used to express gratitude, but more commonly used to alert other drivers to pay attention, there may be obstacles or danger. |

### B. Special Headlight Behavior in Different Countries

TABLE II. SPECIAL HEADLIGHT BEHAVIOR

| Country | Special Headlight Behavior |
|---|---|
| Brazil | Turn signals are not widely used for lane changes or turns. |
| Canada | All hazard lights may indicate parking or encountering emergencies in some areas. |
| Germany | Red triangle on car window and headlights are used as a warning in emergency situations. |
| Italy | High beams may be flashed as a gesture of gratitude or greeting in certain regions. |
| Russia | Drivers often express dissatisfaction or warn others by frequently using flashing lights during traffic congestion. |
| Sweden | Headlights are required during daytime driving, even in good weather conditions. |
| India | Car lights are used to indicate intentions, such as at intersections without traffic signals. |
| Philippines | There are varied interpretations of headlight signals due to common interactions between drivers. |
| South Korea | There is less prevalent use of turn signals for lane changes or turns on highways. |
| Australia | Two flashes of high beams may signal the presence of police nearby. |
| Arab countries | Hazard lights may be flashed to express gratitude or invite other drivers to pass. |
| South Africa | Interior lights at night are used to greet or warn pedestrians or other drivers. |

In addition to the significant variation in interpretation the use of double flashing, there are also notable instances of unique and uncommon headlight behaviors observed in different countries. These behaviors reflect distinct cultural norms and practices, shaping road communication in unconventional ways. For instance, in Brazil, the widespread use of turn signals for lane changes or turns is not common, presenting a deviation from conventional road practices. In Russia, drivers resort to frequent use of horns and flashing lights during traffic congestion to express dissatisfaction or warn others, a behavior seldom seen elsewhere. In the Philippines, interactions between drivers are frequent, leading to varied interpretations of headlight signals amidst the use of car lights, horns, and gestures for communication. Similarly, in Sweden, the requirement to have headlights on during daytime driving, even in favorable weather conditions, stands out as a unique road safety measure [9], [10]. To provide a comprehensive overview, Table II illustrates the cultural diversity in special headlight behavior observed across different countries, emphasizing the necessity of understanding these differences for effective road communication and safer driving practices.

Section II highlights the significance of lighting signals in driving behaviors and their cross-cultural variances. These signals serve as vital means of communication on the road but can be interpreted differently across various cultures, reflecting diverse societal norms and expectations. In discussing cultural variances in double flashing lights behavior and special headlight behaviors, it is observed that how car light signals convey different meanings based on cultural norms and practices. Understanding these nuances depicted in Tables I and II is essential for promoting effective road communication and safer driving practices globally.

## III. DRIVER GESTURES

Human gestures play a vital role in communication on the road, conveying messages of politeness, urgency, and frustration. However, the meaning and appropriateness of gestures vary across cultures, reflecting cultural norms and social expectations [11].

In the realm of driving, driver gestures constitute a vital aspect of road communication, yet their interpretation can vary significantly across different cultures, reflecting diverse societal norms and expectations. In certain regions, gestures such as waving or nodding may convey expressions of gratitude or acknowledgment, while in others, they may be interpreted differently or even deemed offensive [12], [13]. For instance, a thumbs-up gesture commonly signifies approval or agreement in Western cultures, but it may be considered rude or offensive in certain Eastern cultures [13]. Similarly, hand signals used to indicate turns or lane changes may vary in their form and interpretation across different regions. Understanding these cultural nuances in driver gestures is crucial for effective communication on the road and avoiding misunderstandings or conflicts.

The influence of cultural differences on road communication is further exemplified by unique and region-specific driver gestures. For example, in Italy, drivers may employ a hand gesture known as the "Italian Salute" to convey frustration or annoyance, while in Japan, a hand wave known as the "shaka" is used to express gratitude towards other drivers or pedestrians [14]. In India, the head bobble is a prevalent gesture that can indicate agreement, acknowledgment, or even disagreement depending on its context and speed [15]. Moreover, certain cultures may have specific gestures to denote caution, urgency, or appreciation, which may not be universally understood. Recognizing and respecting these unique driver gestures is essential for fostering harmonious road interactions and promoting safer driving practices globally. To give a comprehensive worldwide overview, Table III presents a summary of common and unique driver gestures across different countries, highlighting the cultural variances and distinctive practices observed in road communication.

TABLE III. SPECIAL DRIVER GESTURES

| Country | Special Driver Gestures Across Cultures |
|---|---|
| Brazil | Waving hand out the window to signal acknowledgment |
| Canada | Raised palm to indicate yielding or stopping |
| United States | Originating in Hawaii, the "Hang Loose" hand gesture for relaxation or chill vibes |
| France | Raised hand to thank or acknowledge another driver |
| Germany [16] | Beckoning hand gesture for inviting another driver to proceed |
| Italy | "Italian Salute" hand gesture for frustration or annoyance |
| China | Nodding or hand wave to signal agreement or acknowledgment |
| India | Head bobble for agreement or acknowledgment |
| Japan | "Shaka" hand wave for thanks |
| South Korea [17] | Bowing gesture as a sign of respect or acknowledgment |
| Australia | Raised index finger to indicate caution or warning |
| Saudi Arabia [18] | Hand to the chest as a sign of respect or acknowledgment |
| South Africa | Hand wave to signal thanks or acknowledgment |

As illustrated in Table III, across diverse cultural contexts, drivers utilize various gestures to communicate with fellow road users. For instance, a basic hand wave may express gratitude or acknowledgment in certain cultures, whereas in others, it might be construed as an indication to proceed. Similarly, gestures like finger-pointing or head nodding may convey different meanings based on cultural norms. In summary, Section III underscores the pivotal role of cultural disparities in driver gestures, underscoring the necessity of comprehending and respecting these distinctions for efficient road communication and enhanced safety while driving in cross-cultural settings.

## IV. HONKING

The honking behvaiors or habits across various countries are summarized in Table IV. The functions of honkings can be primarily classified into three categories:

*1) Traffic-related*: Serve functional purposes in traffic, such as avoiding potential dangers (e.g., when vehicles are in close proximity, alert pedestrians of a vehicle's presence), facilitating successful manuevers (e.g., lane changes or overtaking), or prompting movement (e.g., the vehicle ahead remain still when the green light on).

*2) Emotion-related*: Express emotions towards other drivers, pedestrians, or cyclists encountered while driving. They often convey complaints or anger regarding illegal driving behaviors, violations of the right of way, or unexpected maneuvers. This usually comes with aggressive driving.

*3) Social interaction*: Integrated into social interactions, serving as part of daily life to convey greetings, expressions of thankfulness, and other social cues etc.

The aforementioned functions of honking are observed across various countries, without a singular existence. However, certain functions are notably more prominent in specific countries. Research suggests that honking behaviors across cultures may be partly explained by a country's language context [19], characterized as high context (HC) or low context (LC) countries. HC communication relies on implicit and non-verbal cues; while LC communication favors explicit and verbal expression. These context differences also shape how individuals comply and interact within their societies. Consequently, the meanings of honking are influenced: honking may be more prevalent and acceptable in HC societies, while LC societies may exhibit a quieter traffic environment.

TABLE IV. FREQUENCY AND FUNCTION OF HONKING

| Country | Frequency | Prominent function |
|---|---|---|
| Canada, United States [20] | Uncommon[a] | Traffic-related |
| Italy [21], Greece | Very common | Traffic-related; Social-norm; |
| Germany, Switzerland, the Netherlands, | Very uncommon, legally used only as a warning signal | Traffic-related |
| Norway, Finland, Iceland, Sweden, Denmark | Very uncommon, legally used only as a warning signal | Traffic-related |
| China [22], [23] | Common | Traffic-related; Emotion-related |
| India [21], [24], [25] | Very common | Traffic-related; Emotion-related; Social-norm |
| Thailand [20], Vietnam | Very common | Traffic-related; Social norm; Emotion-related |
| Egypt [26], [27] | Common | Traffic-related; Social norm |

*Note*: a. Except for big cities such as New York City

As illustrated in Table IV, the frequency of honking usage and its primary function vary significantly across countries and regions. In typical Asian countries, e.g., China, India, and Thailand, honking is frequently utilized, while in most European countries (with exceptions like Italy and Greece) and North America, honking usage is uncommon, it may be even restricted to serving as a warning signal by law.

There are some pioneering studies regarding honking usages. Statistical analyses in the literature involved the following aspects:

1) Honking is frequently studied as a measure of aggressive driving behavior. Surveys and naturalistic studies [28], employing statistical methods (such as relative risks and odds ratios [29]), have identified factors influencing honking. The factors include sociodemographic (e.g., age, gender), traffic flow characteristics (e.g., speed, flow, Level of Service, density) [30], and culture stereotypes (e.g., nationality) [31];

2) Some studies have investigated the environmental impacts of honking, particularly its effects on traffic noise pollution [32];

3) Additionally, research has examined the frequency of honkings during and after specific religious events like Ramadan in Islam [30].

## V. SPEED CONTROL AND CHOICE

Different speed control and choice practices vary significantly across countries, reflecting unique cultural norms and regulatory frameworks. For instance, in China, it is common for drivers on the main road to slow down to let vehicles on the side road merge, even if they have the "right of way." This practice emphasizes courtesy and cooperation among drivers. In contrast, drivers in North America seldom practice this way, and they usually adhere strictly to right-of-way rules, prioritizing legal precedence over courtesy. The differences regarding speed control and choice across various countries, as reported in the available literature, are summarized in Table V.

TABLE V. SPEED CONTROL AND CHOICE

| Country/ Region | Practice, Evidence, and Findings |
|---|---|
| Germany vs. the UK [3] | When merging onto the motorway, in the UK, the vehicle on the slip road is expected to accelerate, followed in second place by decelerating. In Germany, on the other hand, decelerating is clearly the preferred behaviour. |
| Finland, vs. Sweden, vs. Greece, vs. Turkey [6] | 1) Greek and Swedish drivers reported disregarding the speed limit on residential roads less frequently than Turkish and Finnish drivers; 2) Greek and Turkish drivers reported disregarding the speed limit on a motorway the least frequently with Finnish and Swedish drivers sharing 2$^{nd}$ place; 3) Finnish and Swedish drivers reported overtaking slow drivers on the inside the least frequently, with Greek and Turkish drivers ranking 2$^{nd}$ and 3$^{rd}$. |
| Australia, vs. Netherlands, vs. UK [33] | On roads lacking obvious speed limit signs, the visually perceived speed limit influences drivers' speed. Drivers in the Netherlands perceive speed limits to be significantly lower compared to drivers in the UK and Australia. |
| **Speeding and speed penalties** | |
| Sweden vs. Turkey [5] | Swedish reported a more positive attitude towards complying with the speed limit than Turkish |
| Italy, China [7] | Both countries implement relatively severe penalties for speeding and enforce these penalties rigorously. |
| Egypt [7] | The enforcement of speed penalties may not be adequate to ensure their effectiveness. |
| Japan [7] | While penalties for speeding are higher on urban roads compared to motorways, there are fewer enforcement cameras installed on urban roads than on motorways. |
| North America, vs. Europe regions, vs. Asia/Oceania, vs. Africa [34] | Speeding behavior is more common in North America and Europe regions (more than 55% of respondents) compared to Asia/Oceania and Africa |
| Europe [35] | In Cyprus, Slovenia, and the UK, almost 40% of drivers believe they are (very) often or always checked for speeding, whereas in Sweden, Denmark, and Italy, this percentage is around 5%. |

As illustrated in Table V, pioneering studies have explored various aspects of speed control and choice, providing valuable insights into cross-cultural differences in driving behaviors. One significant aspect pertains to speeding behavior and the enforcement of speed penalties. Notably, speeding appears to be more prevalent in North America and Europe compared to Asia/Oceania and Africa. This trend is further corroborated by [36], revealing that participants from Nordic and Germanic cultural clusters exhibit the riskiest attitudes and behaviors toward speeding, while those from African and Asian cultural clusters demonstrate the least risky behaviors and attitudes. Moreover, countries such as China and Italy enforce relatively severe penalties for speeding, contributing to a lower percentage of individuals checked for speeding.

## VI. ROADMAP FOR FUTURE RESEARCH

Based on the findings from the previous sections, significant cross-cultural variations in implicit driving cues, such as lighting signals, gestures, honking, and speed choice and control, have been observed. However, scientific studies analyzing and comparing these differences, particularly regarding lighting signals and gestures, are very limited (with rare publications identified). Furthermore, existing studies predominantly rely on questionnaires, surveys, or simulator-based datasets, highlighting a notable absence of cross-cultural empirical data. There is an imperative need to establish a worldwide cross-continent empirical database collecting real-world data to facilitate future research endeavors in this domain.

Drawing from these observations, a research roadmap is crafted (as illustrated in Figure ) for future research to delve deeper into these cross-cultural disparities. The goal is to enrich intercultural communication, bolster road safety, and enhance worldwide transportation safety and efficiency. To elaborate, the proposed research roadmap encompasses:

(1) Establish cross-cultural research and evaluation committee: Form a consortium comprising experts from various countries, continents, and cultural backgrounds to collaborate on research related to implicit driving cues and cross-cultural differences. This consortium will facilitate knowledge sharing, data exchange, and collaborative research and evaluation efforts.

(2) Select typical use cases: Identify and select key research priorities and corresponding representative driving scenarios and use cases that capture the nuances of implicit driving cues across different cultural backgrounds.

(3) Design a standardized data collection pipeline: Develop a standardized process for collecting data from different countries with diverse cultural backgrounds. This pipeline should ensure consistency and comparability of data across regions while also guaranteeing privacy protection. Subsequently, collect cross-national and cross-cultural data in different countries under the uniform protocol to construct a worldwide database shared with the research communities.

(4) Knowledge discovery, data mining, analysis and comparison: With the collected cross-cultural dataset, utilize data mining techniques to extract valuable insights and compare driving behaviors and cues across cultures. This step involves analyzing the collected data to identify patterns, trends, and differences in implicit driving cues.

(5) Design online uniformed survey: Develop a comprehensive online survey based on the discovered knowledge to gather feedback and insights from various stakeholders, including drivers, researchers, traffic experts, and policymakers.

(6) Implement crowdsourcing strategy: Implement a crowdsourcing strategy to collecta substantial volume of feedback from stakeholders across diverse countries. This entails developing online platforms for anonymously sharing collected driving behavior data (e.g., video clips) together with engaging participants and experts from other various cultures to offer evaluations and insights into the recorded driving behaviors.

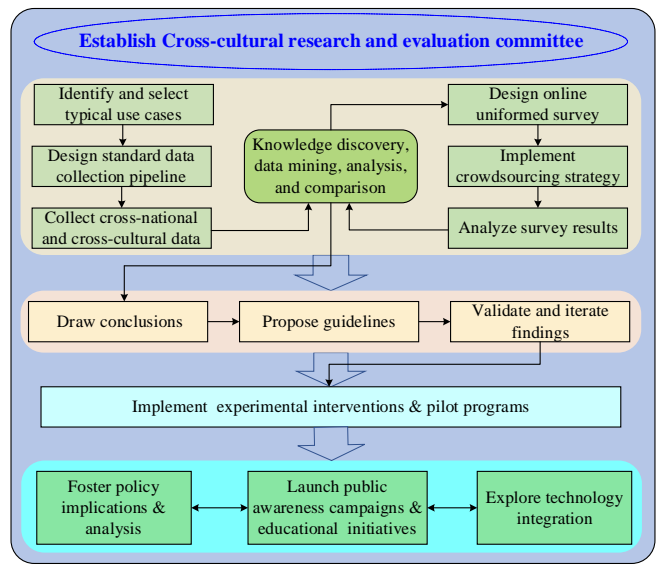

Figure 1.  Roadmap for future research

(7) Analyze survey results: Analyze the survey results to understand stakeholders' perceptions, preferences, and attitudes towards driving behaviors across cultures.

(8) Draw conclusions, propose guidelines, and validate and iterate the findings: Based on the analysis of survey results, data mining, and knowledge discovery, derive conclusions and propose corresponding guidelines for promoting cross-cultural understanding, to foster cross-cultural understanding. Validate and refine the findings through iterative processes to ensure robustness and applicability.

(9) Carry out experimental interventions: Implement experimental interventions and/or pilot programs to test the effectiveness of proposed guidelines and recommendations in real-world driving scenarios. These interventions can provide valuable insights into the practical implications of cross-cultural understanding in improving road safety and transportation efficiency.

(10) Foster policy implications and analysis: Analyze the policy implications of the research findings and propose policy recommendations to policymakers and government agencies. These recommendations may include legislative measures, educational initiatives, and infrastructure improvements aimed at promoting cross-cultural understanding and enhancing road safety.

(11) Launch public awareness campaigns & education initiatives: Initiate public awareness campaigns to raise awareness about the importance of cross-cultural understanding in driving behaviors variation and their impacts on road safety, especially referring to the delivered research findings. Utilize diverse communication channels such as social media, educational workshops, and community outreach programs to disseminate information widely and cultivate cultural sensitivity among drivers.

(12) Explore technology integration: Investigate avenues for integrating technological solutions, such as artificial intelligence, machine learning, and sensor-based systems, to support cross-cultural communication and adaptation in driving environments. These technologies can assist drivers in interpreting implicit cues, navigating diverse driving contexts, and mitigating potential conflicts. The discovery of pertinent findings and insights regarding cross-cultural driving differences will also facilitate the development of customized automated driving features tailored to the specific nuances of various cultural backgrounds in different countries.

Last but not least, it is worth mentioning the importance of fostering multidisciplinary collaboration between researchers, policymakers, industry stakeholders, and community representatives to comprehensively address complex issues related to cross-cultural driving behaviors. This collaborative approach enables the integration of diverse perspectives and expertise, leading to the development of holistic solutions. Furthermore, open-source publishing is essential. Publishing the findings, insights, and proposed guidelines in an open-access format facilitates the dissemination of knowledge and informs stakeholders involved in transportation and road safety initiatives globally.

## VII. CONCLUSION

Understanding and acknowledging implicit driving cues across diverse cultural contexts is imperative for fostering safe and efficient transportation systems at a global scale. While previous pilot studies have explored and compared selected implicit driving cues, they have often been confined to a limited number of countries, thereby lacking a comprehensive worldwide analysis. This study seeks to address this gap by conducting a comprehensive review of existing literature, online resources, and expert insights from various countries. It delves into cross-cultural disparities in driving behaviors, with a specific focus on implicit non-verbal driving cues such as, i.e., hand gestures, signal lighting, honking, and different speed control behaviors. Through rigorous comparative analysis, this research illuminates variations in driving cues across different cultural milieus, considering pertinent factors.

Based on the integrated review findings, this study further proposes a comprehensive research roadmap for future research endeavors aimed at exploring and addressing these disparities. This roadmap is designed to enrich intercultural communication, foster interdisciplinary cooperation, fortify road safety measures, initiate standards, and enhance transportation safety and efficiency on a global scale. By pioneering such efforts, this paper strives to cultivate a nuanced comprehension of implicit driving cues amidst diverse cultural contexts. Moreover, given the crucial necessity for automated driving systems to accommodate cross-cultural nuances, the insights garnered from this study are positioned to serve as a foundational cornerstone in customizing automated driving technologies to meet the varied requirements of different countries with diverse cultural backgrounds.